\documentstyle[12pt]{article}
\def\Journal#1#2#3#4{{#1} {\bf #2}, #3 (#4)}
\def\APNY{\em Ann. Phys. (N.Y.)}
\def\CQG{\em Class. Quantum Grav.}
\def\CMP{\em Commun. Math. Phys.}

\def\PRA{{\em Phys. Rev.} A}
\def\PRD{{\em Phys. Rev.} D}

\def\PRS{{\em Proc. Roy. Soc.} A}

\def\be{\begin{equation}}
\def\ee{\end{equation}}
\def\bea{\begin{eqnarray}}
\def\eea{\end{eqnarray}}
\def\bean{\begin{eqnarray*}}
\def\eean{\end{eqnarray*}}

\def\xip{\xi}
\def\thetap{\theta}
\def\lapse{\lambda}
\def\qocl{q_1^{cl}}
\def\qtcl{q_2^{cl}}
\def\qh{\hat Q}
\def\ph{\hat P}
\def\hh{\hat H}
\def\rab{| \; \alpha , \; \beta \rangle}
\def\lab{\langle \alpha , \; \beta \;  }
\def\one{{\cal I}}
\def\proj{{\cal P}}
\def\eab{e^{-(|\alpha|^2 + |\beta |^2)/2}}
\def\Im'{I_{m'} (2 | \xip |)}
\def\rbyrt{\sqrt{{\tilde r}^2 + 4|\xip|^2}}
\def\rxi{|\; \xip \; \rangle}
\def\lxi{\langle\; \xip \;}
\def\rt{\tilde r}
\def\rtp{{\tilde r}_+}
\def\rtm{{\tilde r}_-}
\def\rtbyr{\sqrt{ r^2 - 4|\xip|^2}}
\def\fm'{f_{m'}(|\xip|)}
\def\Km'{K_{m'} (2 | \xip |)}
\def\rqp{|\; q , \; p \; \rangle}
\def\lqp{\langle q , \; p \; }

\def\opint{\frac{1}{2 \, (2|\xip|)^{m'}\Km'} \ 
\int_{-\infty}^\infty d \rt \ \int_0^{2\pi} \frac{d \thetap}{2\pi} \ 
\frac{e^{-r}}{r} \ \rtp^{m'} \ o(\xip, \rt, \thetap) }
\def\opintr{\frac{1}{2 \, (2|\xip|)^{m'}\Km'} \ 
\int_{-\infty}^\infty d \rt \ 
\frac{e^{-r}}{r} \ \rtp^{m'}}

\renewcommand{\thefootnote}{\fnsymbol{footnote}}

\begin{document}
\bibliographystyle{unsrt}
\hfill MMC-M-12 \\
\hspace*{12.3 cm} September \ 1999 \\
\begin{center}
{\Large  {\bf Coherent state approach to quantum clock in a model where
the Hamiltonian is a difference  between two harmonic oscillators}} 
\\ \vskip 1.2cm
Yoshiaki Ohkuwa \footnote{Email address: ohkuwa@post.miyazaki-med.ac.jp}
\\Department of Mathematics, Miyazaki Medical College,
Kiyotake,\\ Miyazaki 889-1692, Japan \\ \vskip 0.3cm 
\end{center}
\vskip 0.5cm
\begin{abstract}
In order to study the "problem of time", Rovelli proposed a model of a
two harmonic oscillator system where one of the oscillators can be
thought of as a 'clock' for the other oscillator. In this paper we
examine a model where the Hamiltonian is a difference between two
harmonic oscillators, and we consider one of them which has the minus
sign as a 'clock'. Klauder's projection operator approach to generalized
coherent states is used to define physical states and operators. 
The resolution of unity is derived in terms of a gauge invariant
coordinate.  We
investigate the 'quantum clock' and show that the evolution described by
it is identical to the classical motion when the energy becomes large.  
\\

\noindent PACS number(s): 04.60.Ds, 04.60.Kz , 42.50.Ar
\end{abstract}

\renewcommand{\thefootnote}{\arabic{footnote}}

\newpage
\section*{I. INTRODUCTION}

\indent One of the measure conceptual problems in quantum gravity is the
 "problem of time" [1].  
In order to study it Rovelli proposed an interesting model  of a two
harmonic oscillator system where one of the oscillators can be thought of
as a 'clock' for the other oscillator [2]. He showed that the 'clock'
can describe a natural time evolution, even though the system has a time
reparametrization invariance.  
In a similar model Lawrie and Epp studied
an evolution which is governed  by an exact Heisenberg equation, and they
considered coherent states and  introduced  a window function to
investigate an approximate analytical time dependence of the system [3]. 
Recently, Ashworth utilized Klauder's projection operator approach  to
generalized coherent states [4] for the double harmonic oscillator system
[5]. Using Marolf's gauge invariant statement [6], he introduced 'time'
by the phase of an oscillator, 'clock', and he showed that the time
evolution described by the 'clock' agrees with the classical equation of
motion when the energy becomes large.

On the other hand it is well known that the gravitational degree of
freedom has a minus sign in the Hamiltonian of quantum 
cosmology [7].   
The Hamiltonian can be written as a difference between
two harmonic oscillators in some cases, for example, the five-dimensional
Kaluza-Klein cosmology by Wudka [8] and the minisuperspace
model by Hartle-Hawking [9] if time variable is redefined
and the cosmological constant  is assumed to be zero.

In this paper we examine a model where the Hamiltonian is a difference
between two harmonic oscillators, and we consider one of them which has
the minus sign in the Hamiltonian as a 'clock'. The projection operator
approach to generalized coherent states is used to define physical
states.  We deduce a resolution of unity with respect to gauge invariant
states by virtue of a coordinate transformation. In the same way physical
operators are expressed in terms of gauge invariant states and physical
symbols.  We investigate the 'quantum clock' and show that the evolution
described by it is identical to the classical motion when the energy
becomes large.

In \S 2 we will consider a model where the Hamiltonian is a difference 
between two harmonic oscillators, 
and we will use the projection operator approach to generalized
coherent states in order to obtain physical states.
In \S 3 the resolution of unity will be derived in terms of a gauge
invariant coordinate.  
In \S 4 we will
project operators to the physical space, and we will define a 'quantum
clock'  and show that the evolution described by
it is same with the classical motion when the energy becomes large.
We summarize in \S 5.
Appendix A is devoted to derive the gauge transformation of our system.  
In Appendix B it will be shown that our result of the
resolution of unity agrees with that in Ref. [10].

\section*{II. A MODEL OF TWO HARMONIC OSCILLATORS}

Let us consider the following action which is a difference 
between two harmonic oscillators:
\bea
S &= &\int \! dt \, L \ , \\
L &= &\frac{1}{2 \lapse} \, 
\biggl[ \biggl( \frac{d q_1}{d t}\biggr)^2 -
\biggl( \frac{d q_2}{d t}\biggr)^2 \ \biggr]
-\frac{\lapse}{2} \; \bigl[ \, \omega^2 
\bigl( q_1^2 - q_2^2\bigr) - 2E \, \bigr]   \ . \nonumber
\eea
The action (1) has the time reparametrization invariance,  
and the Hamiltonian reads
\be
H \ =\ \lapse \; (H_1 - H_2 - E ) \ , 
\ee
where $H_i = \frac{1}{2} (p_i^2 + \omega^2 q_i^2) \ , \ (i = 1,2 ) \ . $ 
If we define the proper time, 
$\tau = \int_0^t dt' \lapse (t') \ , $
the classical equations of motion for $q_1 , q_2$ are  
$
\ {\ddot q}_i = - \omega^2 q_i \ \ (i = 1 , 2 ) \ , \ 
$
 ${\ddot q}_i = d^2 q_i/d \tau^2\ . \ $  
Therefore, $q_1$ and $q_2$  are ordinary harmonic oscillators 
with only one exceptional point that  $q_2$  has a minus sign in the 
Hamiltonian (2).

We write the classical solution of this system as 
\be
\qocl = A \; cos ( \omega \tau + \phi_1 ) \ , \qquad 
\qtcl = B \; cos (- \omega \tau + \phi_2 ) \ , \ 
\ee
where we have assumed that the two harmonic oscillators have 
opposite dependence on the proper time. 
The reason of this assumption is 
because under the gauge transformation, that is the time translation, 
the phases of the two harmonic oscillators are transformed into 
opposite direction, which is discussed 
in Appendix A.\footnote[1]{It was
pointed out by Prof. T. Kubota that the phases of two harmonic oscillators
are transformed into opposite direction under the gauge transformation.} 
Then the classical motion of each
harmonic oscillator can be also written  by another harmonic oscillator
as 
\bea
\qocl &= &A \; cos \, \biggl(- cos^{-1}\frac{\qtcl}{B} + 
\phi_1 + \phi_2 \biggr) \ , \nonumber \\
\qtcl &= &B \; cos \, \biggl(- cos^{-1}\frac{\qocl}{A} + 
\phi_1 + \phi_2 \biggr) \ . 
\eea
This expression shows that either $\qocl$ or $\qtcl$ can be used for a
classical clock for $\qtcl$  or $\qocl$  , respectively.

Since the action (1) has no time derivative of $\lapse$, 
we get the constraint, 
\be
H = 0 \ , \quad {\rm namely}\quad H_1 - H_2 = E \ , 
\ee
which corresponds to the time reparametrization invariance. 
The equations (3), (5) mean that the classical amplitudes of 
the oscillators must satisfy 
\be
(A \omega )^2 - (B \omega)^2 = 2E \ . 
\ee
The consideration in Appendix A suggests that the gauge transformation
can be written as 
$$
\lapse \rightarrow \lapse + \epsilon \ , \quad 
\phi_1 \rightarrow \phi_1 - \epsilon \omega t \ , \quad 
\phi_2 \rightarrow \phi_2 + \epsilon \omega t \ . 
$$
Hence in our case the summation of initial phases of the two harmonic 
oscillators  $\,\phi_1 + \phi_2$ is gauge invariant.

To quantize this model, we impose the canonical commutation relations 
for Heisenberg operators, $\qh_j , \ph_k$  ,
\bea
\Bigl[ \ \qh_j \ , \ \ph_k \ \Bigr] &= &i \;\hbar \; \delta_{jk} \ , \\
\Bigl[ \ \qh_j \ , \ \qh_k \ \Bigr] &= 
&\Bigl[ \ \ph_j \ , \ \ph_k \ \Bigr] \ = \ 0  \ . \nonumber
\eea
Provided we define annihilation operators by
\be
a = \sqrt{\frac{\omega}{2 \hbar}} \; \qh_1 + 
\frac{i}{\sqrt{2 \hbar \omega }} \; \ph_1 \ , \qquad
b = \sqrt{\frac{\omega}{2 \hbar}} \; \qh_2 + 
\frac{i}{\sqrt{2 \hbar \omega }} \; \ph_2 \ , 
\ee 
then we obtain 
$
\, [ a , a^\dagger ] = [ b , b^\dagger ] = 1 \, , \  
[ a , b ] = [ a^\dagger , b^\dagger ] = 0 \, . 
$ 
From Eqs. (5) and (8) the constraint operator can be written as
\be
{\hat \Phi} = a^\dagger a - b^\dagger b - E' \ , 
\ee 
with $E' = E/  \hbar \omega$ .

We start from the the coherent states for the two harmonic oscillators,  
\be
\rab \ = \ 
\eab \; 
\sum_{n,m=0}^\infty \frac{\alpha^n \beta^m }{\sqrt{n!}\sqrt{m!}}\ 
 | n,m \rangle \ , 
\ee
where
$\ | n,m \rangle = \frac{1}{\sqrt{n!}} (a^\dagger)^n
\frac{1}{\sqrt{m!}} (b^\dagger)^m | 0,0 \rangle \ $
 and $\, \alpha , \beta\, $ are arbitrary complex numbers [11]. 
These coherent states satisfy the properties:
\bea
a \rab &= &\alpha \rab \ , \qquad b \rab \ = \  \beta \rab \ , \ \\
\lab \rab &= &1 \ , \qquad  \qquad\quad 
\one = \int\! \frac{d^2 \alpha}{\pi} \frac{d^2 \beta}{\pi}
\rab \lab | \ , \ \nonumber
\eea
with
$d^2 \alpha = d ( Re \; \alpha) \, d ( Im \; \alpha) $ . 
Using Eqs. (8), (11), we obtain the diagonal element of 
$\qh_1$ and $\qh_2$ as 
\bea
q_1 (\alpha , \beta) &= &\lab| \qh_1 \rab \ = \
\sqrt{\frac{\hbar}{2\omega}}\ (\alpha + {\bar \alpha} ) \ ,  \nonumber\\
q_2 (\alpha , \beta) &= &\lab| \qh_2 \rab \ = \
\sqrt{\frac{\hbar}{2\omega}}\ (\beta + {\bar \beta} ) \ . \ 
\eea

In the same way as Ref. [5], we utilize Klauder's projection operator
approach to generalized coherent states [4]. Projecting $\rab$ on the
physical  states as 
\be
\rab_{phys} = \proj \rab \ , \qquad \proj = \int\! d\mu (\lambda) \ 
e^{\; -i \lambda {\hat \Phi}} \ , \ 
\ee
we have 
\bea
\rab_{phys} &= &\eab
\sum_{n,m=0}^\infty \frac{\alpha^n \beta^m}{\sqrt{n!}\sqrt{m!}}\ 
 \int\! d\mu (\lambda) \ 
e^{\; -i \lambda (n-m-E')}\; 
| n,m \rangle \ ,  \nonumber \\[0.5cm]
&= &\eab \; 
\sum_{m=0}^\infty \frac{\alpha^{m+m'} \beta^m}{\sqrt{(m+m')!}\sqrt{m!}}\ 
 | m+m',m \rangle \ . 
\eea
Here we have set  $E' = m' = n-m $ . 
The norms of these states are
\bea
 _{phys}\lab \rab_{phys} &= &\lab | \proj \rab \ , \ \nonumber \\[0.5cm]
&= &e^{-(|\alpha|^2 + |\beta |^2)} \; 
\sum_{m=0}^\infty \frac{|\alpha|^{2(m+m')}|\beta|^{2m}}{m!(m+m')!}
 \ , \ \nonumber \\[0.5cm]
&= &e^{-(|\alpha|^2 + |\beta |^2)} \; \biggl| \frac{\alpha}{\beta}
\biggr|^{m'}  I_{m'} (2 |\alpha \beta |) \ , \ 
\eea
where we have used the formula [12],
\be
\sum_{n=0}^\infty\ \frac{\Bigl(\frac{x}{2}\Bigr)^{2n}}{n!\; (k+n)!} \ 
= \ \biggl(\frac{2}{x} \biggr)^k \; I_k \, (x) \ , 
\ee
and $I_k \, (x)$  is a modified Bessel function. 
Normalized physical states, $\rab_{n.phys}$  can be written as
\bea
\rab_{n.phys} \ &= \ &\rab_{phys} \  / \ \sqrt{ _{phys}\lab \rab_{phys}}
\ , \nonumber \\[0.5cm]
&= & \frac{\Bigl| \frac{\beta}{\alpha} \Bigr|^{m'/2} \alpha^{m'}}
{\sqrt{I_{m'} (2 |\alpha \beta |) }}\ 
\sum_{m=0}^\infty \frac{(\alpha \beta)^m}{\sqrt{(m+m')!}\sqrt{m!}}\  
 | m+m',m \rangle \ . 
\eea

\section*{III. RESOLUTION OF UNITY}
As indicated in Appendix A the gauge transformation generated by the
constraint transforms the complex coordinates as 
$\alpha \rightarrow \alpha e^{i \varphi}$ and 
$\beta \rightarrow \beta e^{-i \varphi}$ . 
Whence, if we define a complex coordinate, 
$
\xip = \alpha \beta \ , 
$
then $\xip$ is gauge independent. Now $|\, \xip|$ is a product of 
the amplitudes of the two harmonic oscillators, 
and $\; {\rm arg} \, \xip\;$ is the sum of their phases.  
We will see that $\xip$ is sufficient to describe the resolution of
unity in the physical space.  
Let us define the minus of the phase of the second
harmonic oscillator $q_2$ that has the minus sign in the Hamiltonian (2)
as
$\thetap$ , that is 
$
\ \beta = |\beta|\; e^{-i \thetap}\ \ (0 \leq \thetap < 2\pi  ) \ . \ 
$ 
In principle, any of the two oscillators could be used as a 'clock'. 
However, we will consider the classical limit as when the energy becomes
large, namely the first oscillator $q_1$ becomes large. 
So we will later regard $q_2$ as a 'clock' and $\thetap$ as 'time' in this
system.  
We can factor out the dependence on
$\thetap$   from $\rab_{n.phys}$ , 
\bea
\rab_{n.phys} &= &e^{i m' \; \thetap} \ \rxi \ , \\[0.5cm]
\rxi &= 
&\frac{\xip^{m'}}{|\xip|^{m'/2}\sqrt{I_{m'}(|2\xip|)}}
\sum_{m=0}^\infty \frac{\xip^m}{\sqrt{(m+m')!}\sqrt{m!}}\  
 | m+m',m \rangle \ . \nonumber
\eea

The unity operator in the full phase space can be projected on the
physical phase space: 
\be
\one' \ = \ \proj \one \proj  \ = \  
\int\! \frac{d^2 \alpha}{\pi} \frac{d^2 \beta}{\pi} \ \proj
\rab \lab | \; \proj \ . 
\ee
Suppose we change the coordinates, 
\bea
{\tilde r} &= &|\alpha |^2 - | \beta |^2 \ , \nonumber\\
e^{-i \thetap} &= &\frac{\beta}{| \beta |} \ , \\
\xip &= &\alpha \beta \ , \nonumber 
\eea
we have
$$
\alpha = \sqrt{\frac{\rtp}{2}}\; \frac{\xip}{| \xip |}\; e^{i \thetap}\ , 
\qquad\beta = \sqrt{\frac{-\rtm}{2}}\;  e^{-i\thetap}\ ,
$$
\vskip 0.3cm
\noindent 
where 
$
\ {\tilde r}_\pm = {\tilde r} \ \pm  \rbyrt \ \ ,
\quad  
\rtp\rtm = -4|\xip|^2 \ . \ 
$
The absolute value of the Jacobian $|J|$ associated with this change of
coordinates is calculated as\footnote[2]{It is easier to consider the
inverse change of coordinates and to derive $|J^{-1}| = 2r$ than to
calculate $|J|$ directly, which was suggested by Prof. T. Kubota.}
\be
|J| \ = \  \frac{1}{2\rbyrt}\ = \ \frac{1}{2r} \ , 
\ee
where we have defined 
$
r = |\alpha |^2 + | \beta |^2 = \rbyrt \ , \ 
(r \geq 2|\xip|) \ . \ 
$
Using Eqs. (15), (18)-(21), we deduce the resolution of unity, 
\bea
\one'  &=  &\int\! \frac{d^2 \alpha}{\pi} \frac{d^2 \beta}{\pi}\; 
| \, \lab | \proj \rab |\ \rxi\lxi | \ , \nonumber \\[0.5cm]
&= &\int\! d{\tilde r}  d \thetap d^2 \xip
\frac{1}{2\pi^2 r}\; e^{-r}
\biggl(\frac{\rtp}{2 |\xip |} \biggr)^{m'} \Im' \ \rxi\lxi | \ . 
\nonumber
\eea
Because 
$
\rt = \pm \rtbyr \quad 
(+ \ {\rm for}\ |\alpha| \geq |\beta| \ , \ 
-\ {\rm for}\ |\alpha| < |\beta|\  ) \ , \ 
$
 we are led to 
\vskip 0.1cm
\bea
\one'  &=  &\int\! \frac{d^2 \xip}{\pi} \ 
\frac{\Im'}{(2|\xip|)^{m'}} \; \fm' \ \rxi\lxi | \ , \\[0.5cm]
\fm' &= &\int_{-\infty}^\infty d\rt\ \frac{e^{-r}}{r} (\rtp)^{m'} \ = \ 
\int_{2|\xip|}^\infty dr\; \frac{e^{-r}}{\rtbyr} \; 
[\ r_+^{m'}\ + \ r_-^{m'}\ ] \ , \nonumber
\eea 
with
$
r_\pm = r \pm \rtbyr \ . \ 
$
Owing to the following formula [13]:
\bea
\int_a^\infty dx \ \frac{(x+\sqrt{x^2-a^2})^\nu \ 
+ \ (x-\sqrt{x^2-a^2})^\nu}{\sqrt{x^2-a^2}} \ e^{-px}  &=  \ 
2 a^\nu K_\nu (ap) \nonumber \\
(\ a>0 \ ,  &Re \; p >0 \ ) \ , \ \nonumber
\eea
where
$K_\nu$
is a modified Bessel function, we can obtain the explicit expression
of 
$\fm'$ as 
\be
\fm' \ = \ 2 \, (2|\xip|)^{m'}\, \Km' \ . \ 
\ee
Finally we can derive the resolution of unity,  
\be
\one' = \ \frac{2}{\pi} \int\! d^2 \xip \ 
\Im' \ \Km' \ \rxi\lxi | \ , \ 
\ee
from Eqs. (22), (23). 

Now the constraint equations (6), (9) suggest that the underlying symmetry
of our model is $SU(1,1)$ . As shown in Appendix B, it is possible to
prove that our result (24) agrees with the equation (3.22) in Ref. [10]
which is the resolution of unity for generalized coherent states
associated with the Lie algebra of  
$SU(1,1)\ . $

\section*{IV. PROJECTION OF OPERATORS AND QUANTUM CLOCK }
According to Ref. [5], let us define a symbol for an arbitrary operator 
${\tilde O}(\qh, \ph)$ on the physical space as 
\be
o(q, p) |_{phys}\ = \ 
\frac{\lqp| \ \proj\ {\tilde O}(\qh, \ph)\ \proj\rqp}{|\lqp|\ \proj\
\rqp|} \ , \ 
\ee
and let us project ${\tilde O}(\qh, \ph)$ to a
well-defined operator on the physical states as 
\be
{\tilde O}(\qh, \ph) |_{phys} = \int\! d\mu(q,p) \ o(q, p)\ 
\proj\rqp\lqp| \proj\ . \ 
\ee
In the same way as the resolution of unity, we can rewrite this equation
into the form,  
\bea
{\tilde O}(\qh, \ph) |_{phys} &= 
&\int\! \frac{d^2 \alpha}{\pi} \frac{d^2 \beta}{\pi} \ 
o(\alpha, \beta)\ 
\proj\rab\lab| \proj\ , \ \nonumber \\[0.5cm]
&= &\frac{2}{\pi} \int\! d^2 \xip \ 
\Im' \; \Km' \ o'(\xip) \ \rxi\lxi| \ , \  \\[0.5cm]
o'(\xip) &= &\opint  \ , \ \nonumber
\eea
where $r$ and $\rtp$ were defined in Eqs. (20), (21). Note that 
$o'(\xip) $ is the projected symbol and $o'(\xip) = 1$  when 
$o(\alpha, \beta) = 1$ .

Unless the symbol $ o(\xip, \rt, \thetap) $ changes very much with respect
to $\rt$ , the integrand of $o'(\xip)$ ($X$ below) approaches a Gaussian
function around  $\rt
\approx m'$ ,  when the energy of the system $E' = m'$ becomes large: 
\bea
X &= &\frac{e^{-\rbyrt}}{2\Km'\rbyrt} \; 
\left(\frac{\rt + \rbyrt}{2|\xip|} \right)^{m'} \ , \nonumber \\[0.5cm]
&\rightarrow &\frac{1}{\sqrt{2\pi m'}}\  {\rm exp}  
\Biggl[- \frac{(\rt - m')^2}{2m'} \Biggr] \ . \ 
\eea
Here we have used the asymptotic form of $K_\nu$ in Ref. [14], 
\bea
K_\nu (\nu z) &\sim &\sqrt{\frac{\pi}{2\nu}}\ 
\frac{e^{-\nu \eta}}{(1 + z^2)^{1/4}} \qquad\qquad 
(\nu \rightarrow \infty ) \ , \\[0.5cm]
\eta &= &\sqrt{1 + z^2}\  + \ 
{\rm log}\ \frac{z}{1 + \sqrt{1 + z^2}} \ ,  \nonumber
\eea
and we have assumed 
$\rt \gg | \xip | \ , \ m' \gg 1 \ . \ $
Fig. 1 demonstrates the relation between $X$ and $\rt$ , when 
$m' = 10, 100, 1000$ and $|\xip| = 1$ .
The limit (28) of $X$ means that 
$
\ \lim_{m' \rightarrow \infty} \int_{-\infty}^\infty d \rt X \ 
= \ 1 \ . \ 
$
Thus $X$ becomes a delta function $\delta (\rt - m')$ in the classical
limit. 
This means that, when $m'$ is large, the projection of the symbol
satisfies 
$
o'(\xip) \approx \int_0^{2\pi}\frac{d \thetap}{2\pi}\  
o(\xip, m', \thetap) \ , \ 
$
and, if the symbol $o$ is gauge independent, 
namely $o$ does not depend on $\thetap$ , we have 
$
o'(\xip) \approx o(\xip, m', \thetap_0) \ , \ 
$
where $\thetap_0$ is an arbitrary constant 
$(0\leq\thetap_0<2\pi)$ . 
\vskip 1cm

\begin{center}
Fig.\,1
\end{center}
\vskip 1cm

For example, let us take $\qh_1$ and $\qh_2$ for ${\tilde O}(\qh, \ph) $
, then we have 
\bea
q'_1(\xip) &\propto &\int_0^{2\pi} d \thetap \ 
{\rm cos} (\phi_+ + \thetap ) \ = \ 0 \ , \ \nonumber\\
q'_2(\xip) &\propto &\int_0^{2\pi} d \thetap \ 
{\rm cos}\, \thetap  \ = \ 0 \ , \nonumber 
\eea
where we have used Eqs. (12), (27) and have defined  
$
\ \xip \ = \ | \xip | \ e^{i\phi_+} \ . \ 
$
This result is rather natural, since the average positions of operators
over one period  of the oscillator are zero [5]. 
Note that the gauge transformation is 'time translation' in this system,
we must choose a specific time to avoid this result.

Following Ashworth, we use Marolf's gauge
invariant statement [6], 
\be
o|_{q=s} \ = \ \int \! d t \ \frac{d q}{d t}\ \delta [ q(t) - s ]\ o(t) \
.
\ 
\ee
Let us consider the second oscillator $q_2$ which has the minus sign in
the Hamiltonian as a 'clock', and let us regard the minus of its phase 
$\thetap$ as 'time' in our system. So we take 
$
\ q = q_2(\thetap)\ , \ 
s \ = \ B\, {\rm cos}\, (\omega \tau - \phi_2) \ = \ \qtcl \ , \ 
$
and we obtain 
\bea
o|_{q_2=s} &= &\int \! d \thetap \ \frac{d q_2}{d \thetap}\ 
\delta [ q_2(\thetap) - B\; {\rm cos}\; (\omega \tau - \phi_2)  ]\
o(\thetap) \ , \nonumber \\[0.1cm]
&= &\int \! d \thetap \ 
\delta [ \thetap -(\omega \tau - \phi_2)  ]\
o(\thetap) \ = \quad  o(\omega \tau - \phi_2) \ . \nonumber
\eea
This means that we can replace 
$\; o(\xip, \rt, \thetap)|_{q_2=s}\ $ by $\ o(\xip, \rt,
\omega \tau - \phi_2)$ ,  
so Eq. (27) gives 
\bea
o'(\xip , s) &= &o'(\xip) |_{q_2=s} \ , \nonumber\\[0.3cm]
&= &\opint  |_{q_2=s} \ , \nonumber\\[0.5cm]
&= &\opintr \ o(\xip, \rt, \omega \tau - \phi_2) \ . \ 
\eea

Choosing $q_1$ as $o$ , we obtain 
\bea
q'_1(\xip) |_{q_2=s} &= &\opintr \sqrt{\frac{\hbar}{\omega}} \sqrt{\rtp} \
{\rm cos} [\phi_+ + (\omega \tau - \phi_2) ] , \nonumber \\[0.5cm] 
&= &\sqrt{\frac{\hbar}{\omega}}\ \frac{ {\rm cos}
[\phi_+ + (\omega \tau - \phi_2) ] }{2 \, (2|\xip|)^{m'}\Km'} \ 
f_{m' + 1/2} (| \xip |) \ , \nonumber 
\eea
where $\phi_+$ is the phase of $\xip$ , and $\,f_{m' + 1/2} (| \xip |)\,
$  is defined in Eqs. (22) with the replacement, $m' \rightarrow m' +
\frac{1}{2} \ . \ $
Since Eq. (23) means that 
$
\, f_{m' + 1/2} (| \xip|)\ = \ 2 \, (2|\xip|)^{m'+1/2}\, 
K_{m' + 1/2} (2 |\xip|) \ , \ 
$
we arrive at 
\be
q'_1(\xip) |_{q_2=s} \ = \ \sqrt{\frac{\hbar}{\omega}} 
\sqrt{2 |\xip|}\ \frac{K_{m' + 1/2}(2|\xip|)}{\Km'}\ 
{\rm cos}\;  ( \, \omega \tau - \phi_2 + \phi_+ \,) \ . \ 
\ee 
In the classical limit $E' = m' \rightarrow \infty$ , 
the asymptotic form of the modified Bessel function (29) gives 
$
\ K_{m' + 1/2}(2|\xip|) / \Km' \approx 
\sqrt{m' / |\xip|} \ , \ 
$
and 
\bea
q'_1(\xip) |_{q_2=s} &\approx &\sqrt{\frac{2\hbar}{\omega}} 
\sqrt{m'}\ 
{\rm cos}\;  ( \, \omega \tau - \phi_2 + \phi_+ \;) \ , \nonumber \\
&\approx &A \ 
{\rm cos}\;  ( \, \omega \tau - \phi_2 + \phi_+ \;) \ .  
\eea 
Here $A$ is the amplitude of the first oscillator, 
and we have used   
$\rt \approx m' \ , \ \rt \gg |\xip| \ . \ $ 
Note that $\xip$ is gauge invariant and its phase $\phi_+$ is the same as
the initial phase sum $\phi_1 + \phi_2 \ . \ $
Hence the right-hand side of Eq. (33) is identical to the classical
solution $\qocl$ in Eqs. (3) . Namely, the evolution of the first
operator $q_1$ described by the 'quantum clock' $q_2$ is identical to the
classical motion when the energy becomes large.

\section*{V. SUMMARY}
We examined a model where the Hamiltonian is a difference
between two harmonic oscillators, and we considered one of them which has
the minus sign in the Hamiltonian as a 'clock'. The projection operator
approach to generalized coherent states was used to define physical
states.  We deduced a resolution of unity with respect to gauge invariant
states . In the same way, physical
operators were expressed in terms of gauge invariant states and physical
symbols.  We investigated the 'quantum clock' and showed that the
evolution described by it is identical to the classical motion when the
energy becomes large.

As a future work, it will be interesting to apply the projection operator
approach to coherent states in order to study  the time evolution of the
five-dimensional Kaluza-Klein cosmology by Wudka [8] and the
minisuperspace model by Hartle-Hawking [9] when the
cosmological constant  is  zero.

\vskip 1.5cm

\section*{ACKNOWLEDGMENTS}
The author would like to thank Prof. T. Kubota  
for valuable suggestions and encouragement.

\vskip 1.5cm

\section*{APPENDIX A}
First we begin by a pair of creation and annihilation operators of a
harmonic oscillator $a^\dagger , a\,$  which satisfy 
$[ a , a^\dagger ] = 1 \ . \ $
Let us formally define the polar decomposition of $\ a\ $ as
$$
a = e^{i\theta_a}\ | a| \ , \eqno{(A1)} 
$$
where  $| a |$ and $\theta_a$ are the absolute value operator and the
phase operator of $\ a$ , respectively [15].  
Then the number operator 
$\ N_a = a^\dagger a = |a|^2 $
satisfies
$$
[ N_a , e^{i \theta_a} ] = - e^{i \theta_a}\ . \eqno{(A2)}
$$
In the following expansion, 
$$
e^{i \theta_a} N_a e^{-i \theta_a}\ = \ 
N_a + i [ \theta_a , N_a ] 
+ \frac{i^2}{2!} [[ \theta_a , [ \theta_a , N_a ]] + \cdots \ ,  
$$
the left-hand side is equal to $N_a + 1$ by Eq. (A2), and the right-hand
side becomes  $N_a + i [ \theta_a , N_a ]\ ,  $ 
because $[ \theta_a , N_a ]$ is a c-number. Therefore we obtain 
$$[ N_a , \theta_a ] = i \ . \eqno{(A3)} $$

Next let us consider another pair of creation and annihilation operators 
$b^\dagger , b$ with $[ b , b^\dagger ] = 1 \ , $ then similar equations
as Eqs. (A1)-(A3) hold with respect to $b$. 
We examine two cases where the Hamiltonian
is the sum or the difference of two harmonic oscillators. 

\noindent Case 1: 
$\hh_+ = \hbar \omega ( N_a + N_b - E' )$

Since Eq. (A3) means $[ \hh_+ , \theta_a ] = i\hbar \omega\ $ and 
$\ [ \hh_+ , \theta_b ] = i\hbar \omega$ , we have 
$$
[ \hh_+ , \theta_a + \theta_b ] \ = \ 2 i\hbar \omega\ , \quad
[ \hh_+ , \theta_a - \theta_b ] \ = \ 0 \ . \eqno{(A4)}
$$
Therefore the phase difference $\theta_a - \theta_b$ is gauge invariant, 
and the phase sum $\theta_a + \theta_b$ is not invariant in this case.
This case was investigated in Ref. [5].

\noindent Case 2: 
$\hh_- = \hbar \omega ( N_a - N_b - E' )$

Since Eq. (A3) means $[ \hh_- , \theta_a ] = i\hbar \omega\ $ and 
$\ [ \hh_- , \theta_b ] = - i\hbar \omega$ , we have 
$$
[ \hh_- , \theta_a + \theta_b ] \ = \ 0 \ , \quad
[ \hh_- , \theta_a - \theta_b ] \ = \ 2 i\hbar \omega\ \ . \eqno{(A5)}
$$
Hence the phase sum $\theta_a + \theta_b$ is gauge invariant,  and the phase difference $\theta_a -
\theta_b$ is not invariant in this case.

From Eq. (9), $\hh_- = \hbar \omega {\hat \Phi}$ , and $\hh_-$ is the
generator of the gauge transformation (time translation) of our
system. So Eqs. (A5) suggest that the phase of each harmonic oscillator
transforms into opposite direction under the gauge transformation.

\section*{APPENDIX B}
Barut and Girardello [10] derived the resolution of unity for
generalized coherent states associated with the Lie algebra of  
$SU(1,1) $ as 
$$
\one' = \int\! d \sigma (z) \ | z \rangle \langle z | \ ,
\qquad\qquad\qquad\qquad
\eqno{(B1)}  
$$
$$
\sigma (z) = \sigma (\rho) \ \rho \, d \rho \, d \varphi \ , \qquad
\sigma (\rho) = \frac{4}{\pi \Gamma(-2\Phi)} 
\Bigl(\sqrt{2}\rho \Bigr)^{-2\Phi -1}
K_{1+2\Phi} \Bigl(2\sqrt{2}\rho \Bigr) \ , 
$$
where 
$z = \rho\, e^{i\varphi} , \ |z| = \rho $ and 
$- 2 \Phi - 1 = 0, 1, 2, \cdots . \ $
Here we have corrected an erratum of $\sigma (\rho)$ in Ref. [10], namely 
$K_{1/2 + \Phi}  \rightarrow K_{1+2\Phi}\,$. 
The reason of this is because the formula in Ref. [16] was wrong, and it
should be replaced by 
$$
\int_0^\infty d x \ 2 x^{\alpha +\beta} K_{2(\alpha - \beta)}\,
(2\sqrt{x}) \  x^{s-1} \ 
= \ \Gamma(2\alpha +s) \, \Gamma(2\beta +s) \ .  \eqno{(B2)}
$$
This formula can be proved by the integral expression of 
$K_\nu (z) $ [17], \break
$
K_\nu (z) = \int_0^\infty d t \ {\rm exp}\, (-z \>{\rm cosh}\, t )\ 
{\rm cosh}\, (\nu t ) \ , \ 
$
a change of the order of integration and the following formula 
 [18]:
$$
\int_0^\infty d x \;\frac{{\rm cosh}\, b x}{{\rm cosh}^\nu\, c x} \ = \ 
\frac{2^{\nu-2}}{c \Gamma (\nu)} \ 
\Gamma \biggl(\frac{\nu c + b}{2 c} \biggr) \ 
\Gamma \biggl(\frac{\nu c - b}{2 c} \biggr) \ . \ 
$$
We can also easily assure Eq. (B2) in a special case that 
$\alpha = 1/4 , \ \beta = 0 , \  s = 1 \ , \ $ using 
$
K_{1/2} (z) = \sqrt{\pi/ 2z}\ {\rm exp}\, (-z) \ \ 
$
 [19].

Let us write the normalized state of $|z \rangle$ as 
$|z \rangle_n$ , then we have 
$$
| z \rangle  \langle z | \ = \ \langle z | z \rangle \ \
| z \rangle_n \,_n\langle z | \ , \qquad\qquad\qquad\eqno{(B3)}
$$
$$
\langle z | z \rangle \ = \ \Gamma (-2\Phi) \ 
\sum_{n=0}^\infty \frac{(\sqrt{2} | z | )^{2n}}{n! \Gamma(-2 \Phi +
n)}\    
= \ \frac{m'!}{(\sqrt{2}|z|)^{m'}}\ I_{m'} (2\sqrt{2}|z|) \ , \ 
$$
where we have used Eq. (16) and have identified 
$m' = -2 \Phi - 1 \ . \ $ 
If we write $\xip = \sqrt{2} z$ and 
$| z \rangle_n \,_n\langle z |  = \rxi \lxi| \ , \ $ 
then Eqs. (B1) is identical to Eq. (24) . 
Therefore, we have established that our resolution of unity agrees with
that in Ref. [10] .


\newpage
\section*{Figure Captions}
FIG. 1. The relation between $X$ and $\rt$ , when 
$m' = 10, 100, 1000$ and $|\xip| = 1$ .


\newpage
\begin{thebibliography}{99}
\bibitem{IshamKucharCQGPT}
C.J. Isham, in {\em Integrable Systems, Quantum Groups, and Quantum Field
Theories},  eds. L.A. Ibort and M.A. Rodriguez
(Kluwer, London, 1993), p. 157; 
K.V. Kucha\v r, in {\em Proceedings of the 4th Canadian Conference 
on General Relativity and Relativistic Astrophysics}, 
eds. G. Kunstatter, D.E. Vincent and J.G. Williams
(World Scientific, Singapore, 1992), p. 211.
\bibitem{RoverlliQMWT}  
C. Rovelli, \Journal{\PRD}{42}{2638}{1990}; D {\bf 43}, 442 (1991).
\bibitem{LawrieEpp}
I.D. Lawrie and R.J. Epp, \Journal{\PRD}{53}{7336}{1996}.
\bibitem{Klauder}
J.R. Klauder, \Journal{\APNY}{254}{419}{1997}.
\bibitem{AshworthCSAQC}
M.C. Ashworth,
\Journal{\PRD}{58}{104008}{1998}; \Journal{\PRA}{57}{2357}{1998}.
\bibitem{MarolfQORD}
D. Marolf, \Journal{\CQG}{12}{1199}{1995}. 
\bibitem{HalliwelILQC}
J.J. Halliwell, in {\em Quantum Cosmology and Baby Universes}, eds. 
S. Coleman, J.B. Hartle, T. Piran and S. Weinberg 
(World Scientific, Singapore, 1991), p. 159.  
\bibitem{Wudka}
J. Wudka, \Journal{\PRD}{35}{3255}{1987}.
\bibitem{Hartle-Hawking}
J.B. Hartle and S.W. Hawking, \Journal{\PRD}{28}{2960}{1983}. 
\bibitem{BarutGirardello}
A.O. Barut and L. Girardello, \Journal{\CMP}{21}{41}{1971}. 
\bibitem{KlauderSkagerstam}
J.R. Klauder and B.-S. Skagerstam, {\em Coherent States} 
(World Scientific, Singapore, 1985).  
\bibitem{AbramowitzStegun}
M. Abramowitz and I. Stegun, {\it Handbook of Mathematical Functions} 
(Dover, New York, 1972), p. 375, Eq. 9.6.10.   
\bibitem{PrudnikovI1}
Y. Ohtsuki and Y. Murotani, {\it New Handbook of Mathematical Formulae 
Vol.1 -Elementary Functions-}  (Maruzen, Tokyo, 1991), p. 332, \S 2.3.11,
Eq. 9, in Japanese, translated
 from Russian; 
English translation: A.P. Prudnikov, Yu.A. Brychkov and O.I. 
 Marichev, 
{\it Integrals and Series Vol.1 -Elementary Functions-} 
(Gordon and Breach, New York, 1986).
\bibitem{AbramowitzStegun2}
Ref. [12], p. 378, Eq. 9.7.8. 
\bibitem{DiracQTEAR}
P.A.M. Dirac, \Journal{\PRS}{114}{243}{1927}. 
\bibitem{Eldelyi}
A. Eld\'elyi (ed.), Bateman Project: Vol. I. Integral transformations,
 (McGraw-Hill, New York, 1954) p. 349.  
\bibitem{AbramowitzStegun3}
Ref. [12], p. 376, Eq. 9.6.24. 
\bibitem{PrudnikovI2} 
Ref. [13], p. 353, \S 2,4,4, Eq. 4.  
\bibitem{PrudnikovII}
Y. Ohtsuki and Y. Murotani, {\it New Handbook of Mathematical Formulae 
Vol.2 -Special Functions-}  (Maruzen, Tokyo, 1992), p. 328, \S 2.16,
 in Japanese, translated
 from Russian; 
English translation: A.P. Prudnikov, Yu.A. Brychkov and O.I. 
 Marichev, 
{\it Integrals and Series Vol.2 -Special Functions-} 
(Gordon and Breach, New York, 1986). 
\end{thebibliography}
\end{document}